\documentclass[twocolumn,showpacs,floats,floatfix,superscriptaddress,aps,pra]{revtex4}
\usepackage{amsfonts}
\usepackage{amssymb}
\usepackage{amsmath}
\usepackage{graphicx}
\usepackage{bm}
\usepackage{color}
\usepackage{epsfig}
\usepackage{calc}
\usepackage{ifthen}
\usepackage{subfigure}
\usepackage{graphicx}
\usepackage{epstopdf}
\usepackage{epsfig}

\begin{document}

\title{Coupling-assisted quasi-bound states in the continuum in heterogeneous metasurfaces}
\date{\today }

\begin{abstract}
In this paper, we present a Bound states in the continuum (BIC) metamaterial in heterogeneous structures based on the universal coupled mode theory. 
We find the more general physical parameters to represent BIC, which are the resonant frequencies and corresponding phases of metamaterial structures. 
Therefore, BIC metamaterial comes from the equal value of the resonant frequencies and phases of metamaterial structures which are not only for homogeneous structures. 
Meanwhile if slightly vary one of metamaterial structure's resonant frequency and phase by varying geometry, we can obtain the quasi-BIC instead of broken symmetry of homogeneous structures.
In this paper, we provide the BIC and quasi-BIC with one example of two heterogeneous structures which are cut wire (CW) and Split-Ring Resonator (SRR), to widely extends the metamaterial BIC beyond common sense. Furthermore, we demonstrate the simulation results and experimental results to proof our idea. 
\end{abstract}

\author{Wei Huang}
\affiliation{Guangxi Key Laboratory of Optoelectronic Information Processing, School of Optoelectronic Engineering, Guilin University of Electronic Technology, Guilin 541004, China}

\author{Songyi Liu}
\affiliation{Guangxi Key Laboratory of Optoelectronic Information Processing, School of Optoelectronic Engineering, Guilin University of Electronic Technology, Guilin 541004, China}

\author{Dehui Zeng}
\affiliation{Guangxi Key Laboratory of Optoelectronic Information Processing, School of Optoelectronic Engineering, Guilin University of Electronic Technology, Guilin 541004, China}

\author{Quanlong Yang}
\affiliation{Nonlinear Physics Centre, School of Physics, Australian National University, Canberra, ACT 2601, Australia}

\author{Wentao Zhang}
\affiliation{Guangxi Key Laboratory of Optoelectronic Information Processing, School of Optoelectronic Engineering, Guilin University of Electronic Technology, Guilin 541004, China}

\author{Shan Yin}
\email{syin@guet.edu.cn}
\affiliation{Guangxi Key Laboratory of Optoelectronic Information Processing, School of Optoelectronic Engineering, Guilin University of Electronic Technology, Guilin 541004, China}

\author{Jiaguang Han}
\email{jiaghan@tju.edu.cn}
\affiliation{Guangxi Key Laboratory of Optoelectronic Information Processing, School of Optoelectronic Engineering, Guilin University of Electronic Technology, Guilin 541004, China}
\affiliation{Center for Terahertz Waves and College of Precision Instrument and Optoelectronics Engineering, Tianjin University, Tianjin 3000072, China}

\maketitle


\section{Introduction}
Bound states in the continuum (BIC) is initially proposed in quantum mechanics \cite{Neumann1929}, which trapped or guided modes with their frequencies in the frequency intervals of radiation modes in optical system \cite{Hsu2016, Marinica2008, Plotnik2011, Gao2016}. Based on this idea, most recently, the concept of BIC has already been introduced into the metamaterial structure, which can offer the ultra-high Q resonance \cite{Azzam2018, Koshelev2018, Koshelev2019, Miroshnichenko2010, Kupriianov2019, Abujetas2019, Cong2015, Liang2020}. 
Based on the metamaterial BIC, there are many practical applications for photonic systems, such as lasers \cite{Kodigala2017, Ha2018}, sensors \cite{Liu2017, Romano2018}, high-sensitive medical devices \cite{Khanikaev2013, Tittl2018} and filters \cite{Foley2014}. 
Due to widely applications, the BIC has been attracted many research interests.

The reason of causing metamaterial BIC is that if two lossy fields of metamaterial structures have the same phase and resonant frequency of single metamaterial structure, the far field of two lossy fields have destructive interference with each others. When we slightly vary one of metamaterial structure, the far field of two lossy fields can not destructive interference, but provides high Q Fano-shape resonance instead, so-called broken symmetry of quasi-BIC. 
The most intuitive choice and common sense for BIC metamaterial is made of two uniform structures and quasi-BIC comes from the broken symmetry of homogeneous structures, which can obtain the high Q resonance. 

\begin{figure}[htbp]
	\centering
		\includegraphics[width=0.5\textwidth]{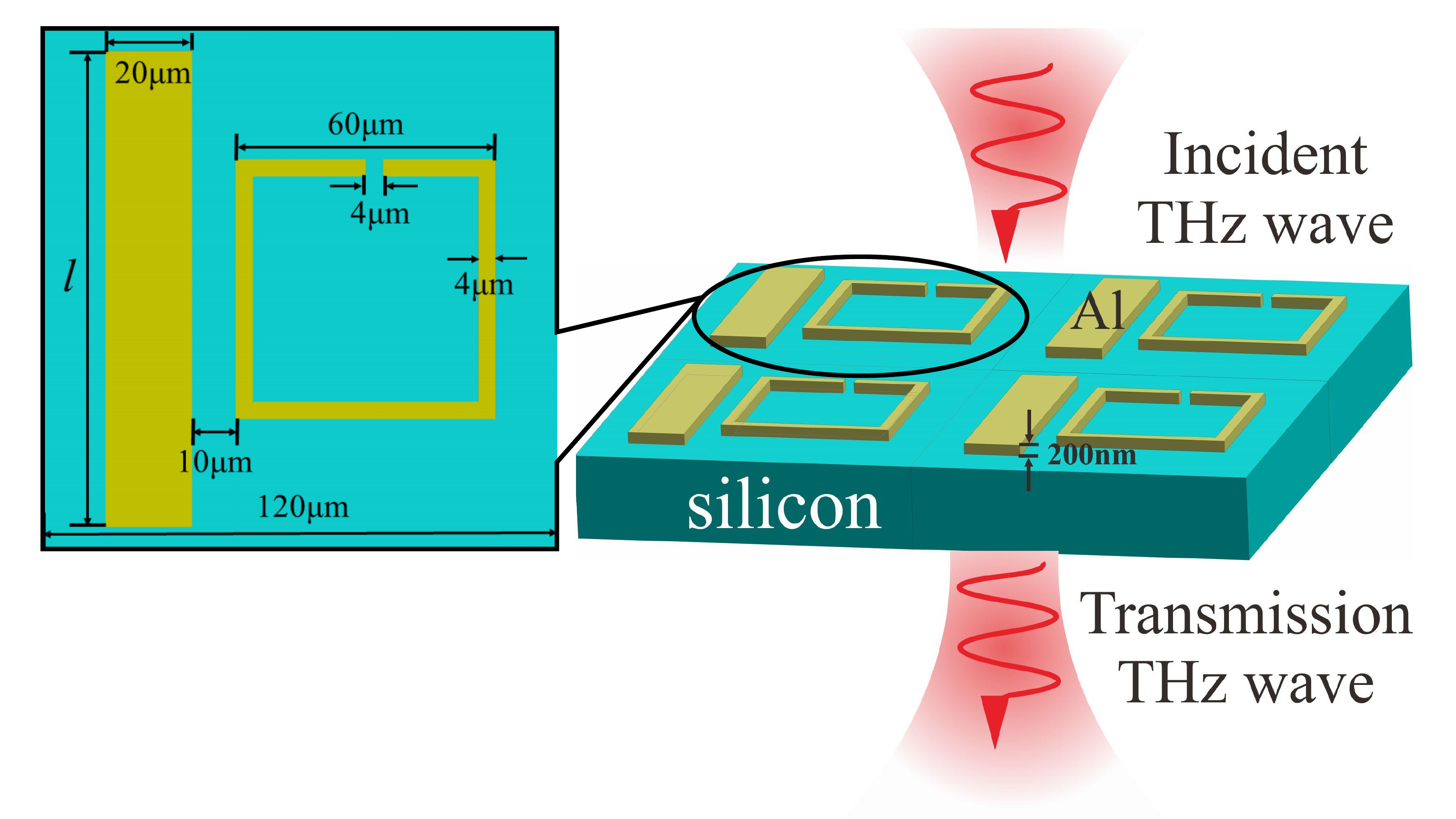}
	\caption{The schematic figure of our new type BIC with two entirely different metamaterial structures (Cut Wire and Split-Ring Resonator). We carefully select the structure of CW and SRR to make the same phase and resonant frequency. }
	\label{Fig1}
\end{figure}

\begin{figure*}[htbp]
	\centering
	\includegraphics[width=1\textwidth]{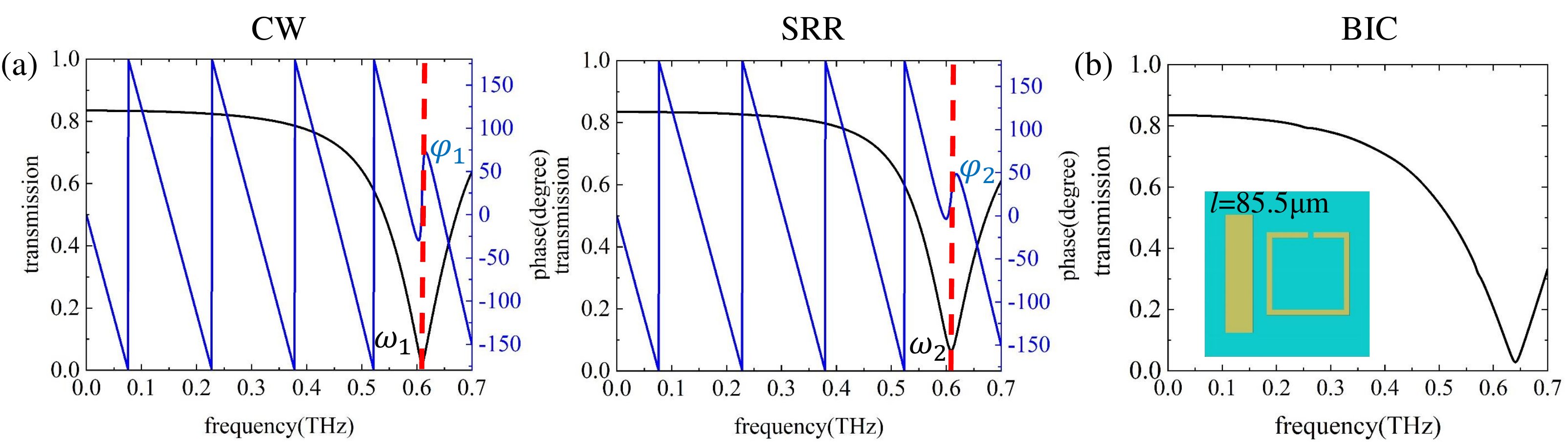}
	\caption{(a)The resonant frequency (upper figures) and corresponding phase at resonant frequency (lower figures) of single metamaterial structure of CW (left figures) and SRR (right figures). The corresponding geometrical parameters of CW are the length $L = 85.5$ $\mu m$, width 20 $\mu m$ and the corresponding geometrical parameters of SRR are gap of SRR is 4 $\mu m$; width of SRR is 4 $\mu m$ and the side length of SRR is 60 $\mu m$, as shown in Fig. \ref{Fig1}. (b) The transmission spectrum of coupling between CW and SRR, which provide the BIC. }
	\label{Fig2}
\end{figure*}

\begin{figure*}[htbp]
	\centering
	\includegraphics[width=1\textwidth]{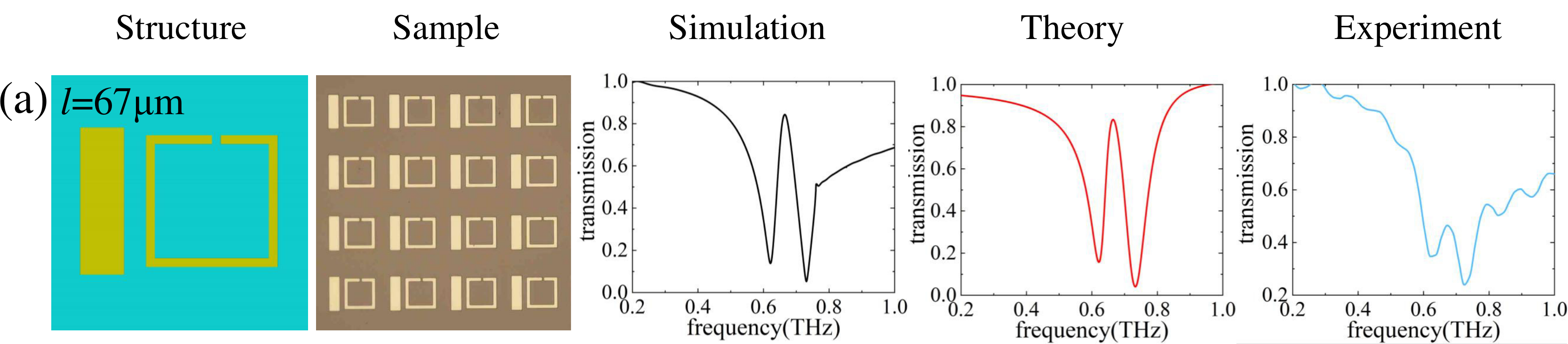}
	\includegraphics[width=1\textwidth]{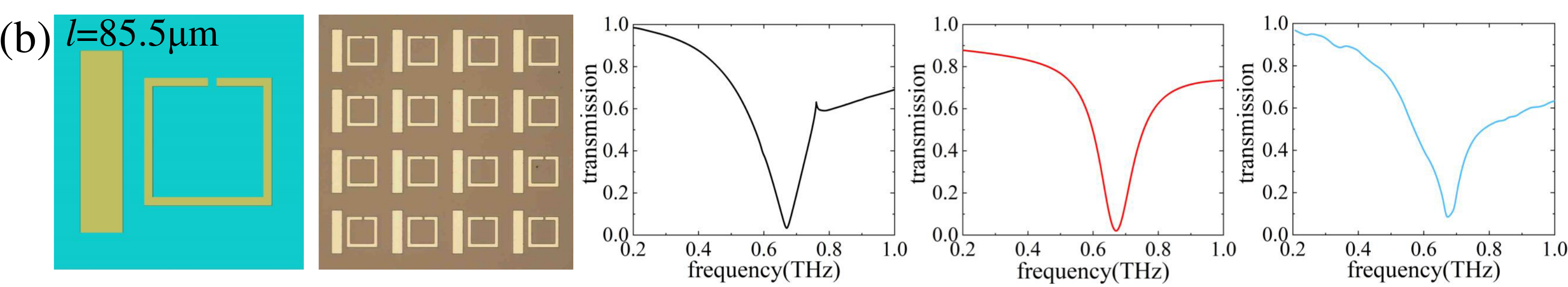}
	\includegraphics[width=1\textwidth]{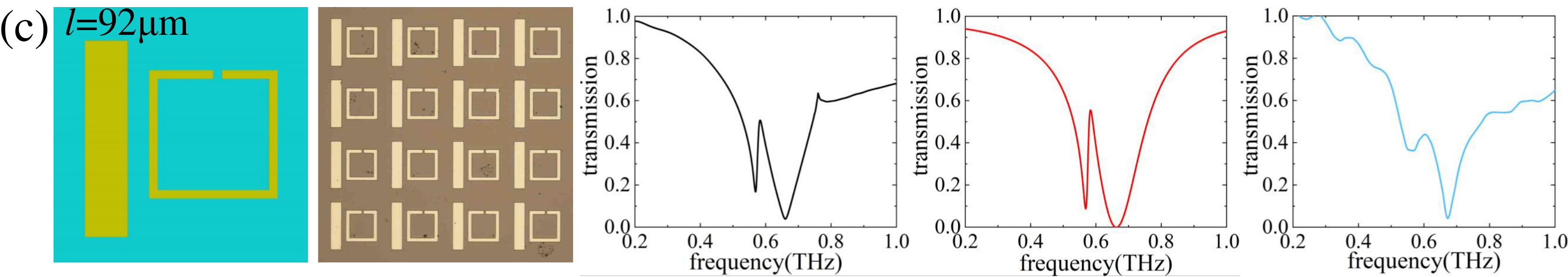}
	\includegraphics[width=1\textwidth]{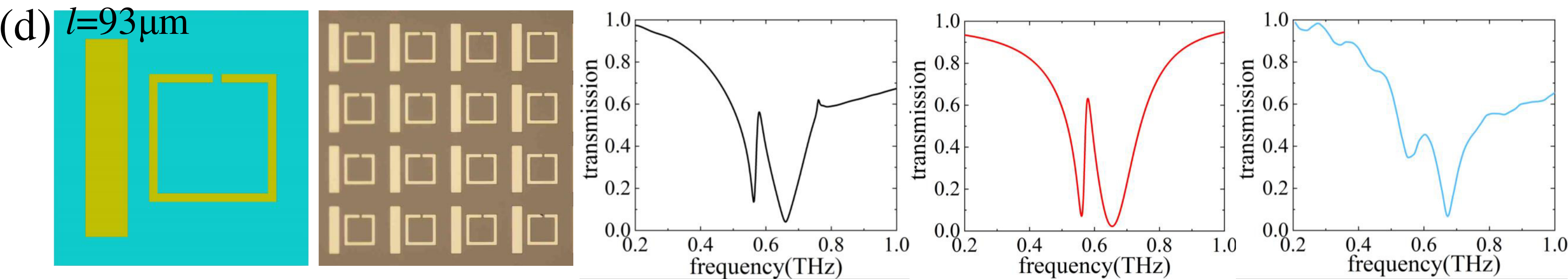}
	\includegraphics[width=1\textwidth]{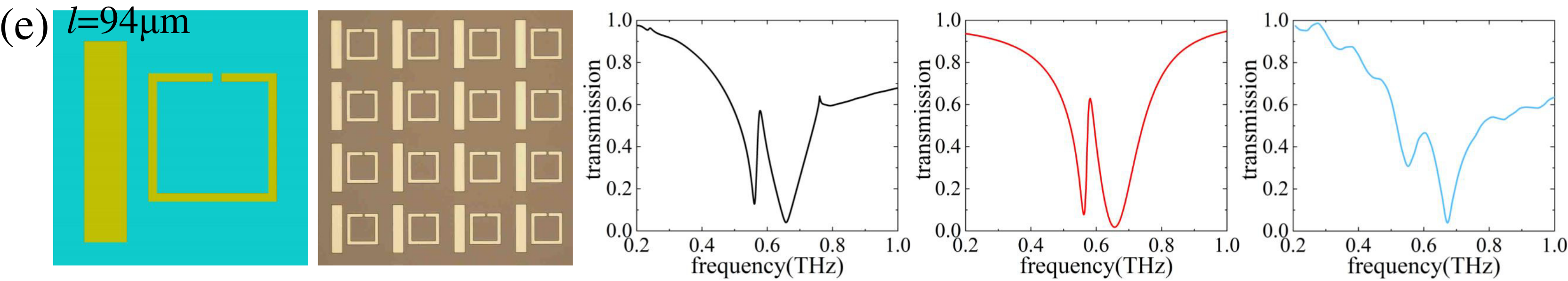}
	\includegraphics[width=1\textwidth]{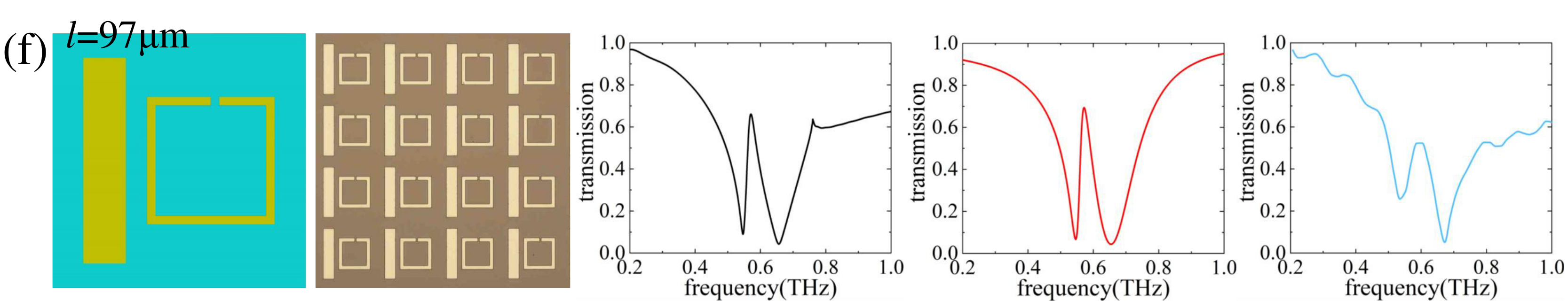}
	\includegraphics[width=1\textwidth]{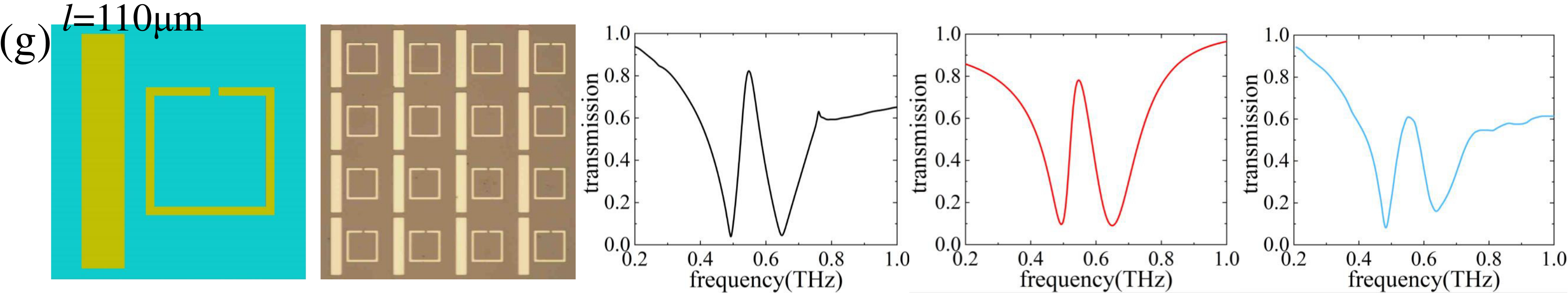}
	\caption{ The transmission spectrum of coupling between CW and SRR with varying the length of CW (a) $L = 67$ $\mu m$, (b) $L = 85.5$ $\mu m$, (c) $L = 92$ $\mu m$, (d) $L = 93$ $\mu m$, (e) $L = 94$ $\mu m$, (f) $L = 97$ $\mu m$, (g) $L = 110$ $\mu m$. }
	\label{Fig3}
\end{figure*}

The coupled mode theory (CMT) is a prosperous and universal theory and it widely used in many systems \cite{Cong2015, Niu2021, Huang2014, Huang2019, Ostrovskaya2000, Huang2020, Koshiba2011, Huang20202, Huang2018}. 
Most recently, CMT implies the more fundamental physical parameters for BIC metamaterial, such as the resonant frequencies and corresponding phases of metamaterial structures \cite{Huang2021}. Therefore, we can simply extend this deduction of establishing BIC metamaterial by employing two heterogeneous structures with the same value of resonant frequencies and corresponding phases. 
In the other words, even for two heterogeneous metamaterial structures (for example, Cut Wire (CW) and Split-Ring Resonator (SRR)) which have the same resonant frequency and corresponding phase, we can obtain the BIC metamaterial with heterogeneous configuration. 
This finding will exceed the common sense of BIC metamaterial with homogeneous structures
Normally, previous researches only realize BIC metamaterial caused by two homogeneous structures (or quasi-BIC caused by two slightly different structures). 
In this paper, we obtain the BIC (and quasi-BIC) with two heterogeneous metamaterial structures, as shown in Fig. \ref{Fig1} (CW and SRR) for the first time. 
We provide the CW and SRR with the same phase and resonant frequency (see Fig. \ref{Fig2}a) for the BIC case as shown in Fig. \ref{Fig2}b. Furthermore, we slightly vary the structure of CW to change the phase and resonant frequency of CW (see Fig. \ref{Fig3}a and  Fig. \ref{Fig3}b), to obtain the corresponding quasi-BIC as shown in Fig. \ref{Fig4}a and Fig. \ref{Fig4}b. Subsequently, we continuously vary the length of CW to find the corresponding Q-value of quasi-BIC, and we give the fitting function according to $\alpha^{-2}$, where $\alpha$ is the asymmetric parameter \cite{Azzam2018}, as shown in Fig.\ref{Fig4}.


The paper is organized as follows. In section II, we firstly introduce the universal coupled mode theory for metamaterial BIC. Section III employs the universal coupled mode theory to predict the BIC with the same phase and resonant frequency of CW and SRR. After that, we slightly change the phase and resonant frequencies of CW, which provides a different Q-value of quasi-BIC. In addition, we give the experimental results to demonstrate our device. 
In the section IV, we provide the further discussions on our design. Finally, we conclude in Section V.

\section{Universal coupled mode theory}
The universal coupled mode theory can be described as following \cite{Huang2021},

\begin{equation}
\left[
	\begin{matrix}
	\ w-w _{a}-i\gamma_{a} & \Omega \\
	\Omega & \ w-w _{b}-i\gamma_{b}%
	\end{matrix}%
	\right] \left[
	\begin{matrix}
	a \\
	b%
	\end{matrix}%
	\right] =\left[
	 \begin{matrix}
	 \sqrt{\gamma_{a}}E \\
	 \sqrt{\gamma_{b}}e^{_{i\phi}}E%
	 \end{matrix}%
	 \right],
\end{equation}%

where $|a|^2$ and $|b|^2$ as the energies in each metamaterial structure. $\omega$ is the frequency of input THz wave. $\omega_a$ and $\omega_b$ represent the frequencies of the metamaterial structures, which is the same as the resonant frequency of the corresponding  metamaterial structure ($\omega_a$ = $\omega_1$; $\omega_b$ = $\omega_b$). $\Omega$ is the coupling strength with loss, due to the loss of transferring energy from one metamaterial structure to another, with $\Omega = g-i\sqrt{\gamma_{a} \gamma_{b}}e^{_{i\phi}}$, where $g$ is the coupling strength between two metamaterial structures. $\gamma_a$ and $\gamma_b$ are the loss terms of metamaterial structures, which is close to $\gamma_1$, $\gamma_2$ for each single metamaterial structure ($\gamma_{a}=\gamma_{1}$;  $\gamma_{b}=(\gamma_{1} - \gamma_{2})/2$). $\phi$ is the phase information, which can be calculated by the phase $\phi_1$ and $\phi_2$ for each metamaterial structure ($\phi=(\phi_{1} - \phi_{2})d$, where $d$ is the width of metamaterial structure). $E$ is the amplitude of external exciting THz wave.

Subsequently, we can obtain the energy amplitudes $a$, $b$ of each metamaterial structure by solving the Eq. 1, as shown,
\begin{equation}
	a =\frac{ ((w-w _{b}-i\gamma_{b}) \sqrt{\gamma_{a}}-\Omega \sqrt{\gamma_{b}}e^{_{i\phi}})E} { ( w-w _{b}-i\gamma_{b}) (w-w _{a}-i\gamma_{a})-\Omega^{2}};
\end{equation}
	
\begin{equation}
	b =\frac{ ((w-w _{a}-i\gamma_{a}) \sqrt{\gamma_{b}}-\Omega \sqrt{\gamma_{a}})E} { ( w-w _{b}-i\gamma_{b}) (w-w _{a}-i\gamma_{a})-\Omega^{2}}.
\end{equation}

Then we calculate the effective susceptibility which is the linear superposition with energy amplitudes $|a|^2$ and $|b|^2$. the effective electric susceptibility of the metamaterial can be written as \cite{meng2012}

\begin{equation}
\chi_{\text{eff}} =\frac{ \sqrt{\gamma_{a}}a+\sqrt{\gamma_{b}}e^{_{i\phi}}b} {\epsilon_{0} E }.
\end{equation}
Finally, we obtain the transmission spectrum with $T \approx 1- \text{Im}(\chi_{\text{eff}})$ \cite{Cong2015} , as shown,
\begin{widetext}
\begin{equation}
	T \approx 1-\text{Im}(\frac{ (w-w _{a}-i\gamma_{a}) \gamma_{b}e^{_{2i\phi}}+((w-w _{b}-i\gamma_{b}) \gamma_{a}-2\Omega \sqrt{\gamma_{a}\gamma_{b}}e^{_{i\phi}}} { ( w-w _{b}-i\gamma_{b}) (w-w _{a}-i\gamma_{a})-\Omega^{2}}).
\end{equation}
\end{widetext}
Hence, we can predict the transmission spectrum of the structure \cite{Huang2021}, to obtain the BIC and Q-value of quasi-BIC. 


Therefore, our theory implies that the BIC only is relative to the phases and resonant frequencies of two adjacent metamaterial structures. Thus, if two metamaterial structures have the same phase and resonant frequency, they should have the BIC phenomenon even for two entirely different structures. 
In other words, if we carefully design two heterogeneous structures which have the same value of resonant frequency and corresponding phase, we can establish BIC metamaterial beyond the common sense of homogeneous structures. Furthermore, when we slightly vary resonant frequency and corresponding phase of any one heterogeneous structures by changing the geometrical parameters, the quasi-BIC with two heterogeneous structures will appear. 
In this paper, we takes the CW and SRR as the example to demonstrate our idea, but our results are not only for CW and SRR.

\section{BIC with two different structures}

Based on the prediction of our theory, we carefully select the structures of CW and SRR to obtain the BIC or quasi-BIC. In other to easier demonstration, we fix the  structure parameters of SRR, where the gap of SRR is 4 $\mu m$; width of SRR is 4 $\mu m$ and the side length of SRR is 60 $\mu m$. Besides, the width of CW is fixed 20 $\mu m$ with the length $L$ of CW, as shown in Fig. \ref{Fig1}. Therefore, we just required to change the length $L$ of CW to vary the phase and resonant frequency of CW, to obtain the BIC or quasi-BIC. 

For the BIC case, we select the length $L_0$ of CW as 85.5 $\mu m$, where the resonant frequency of CW is 0.6097 THz and the corresponding phase at the resonant frequency is 26.125 degrees. Besides, the resonant frequency of SRR is 0.6095 THz and corresponding phase at resonant frequency is 26.051 degrees, as shown in Fig. \ref{Fig2}a. Thus, according to our theory, this configuration generates the ultra-high Q resonance or we can call it as BIC. 
The transmission spectrum of CW and SRR coupling demonstrates in Fig. \ref{Fig2}b. From the results, coupling between two entirely different structures performs the transmission spectrum as the single resonance. Thus, the Q-value at BIC frequency reveals the infinite. 

Based on our universal coupled mode theory, if we slightly change the length $L$ which varies the resonant frequencies and the corresponding phases of CW, the resonant frequencies and corresponding phases of CW and SRR are not match. Therefore, the Fano resonant shapes will appear in the transmission spectrum, so-called quasi-BIC cases. 
In order to demonstrate the correct prediction of our theory, we propose the full-wave simulations, our theory and experiments by varying the length of CW from $L = 67$ $\mu m$ to $110$ $\mu m$, as shown in Fig. \ref{Fig3}. The first column and second column of Fig. \ref{Fig3} demonstrate our designed structure of units cells and our experimental devices, respectively. The third, fourth, and fifth columns are the full-wave simulations, our theory and corresponding experimental results, respectively. 
In Fig. \ref{Fig3}, our theoretical results can be well-fitted to our full-wave simulations. Meanwhile, our experimental results are also well consistent with our theoretical and full-wave simulated results, comparing with the resonance frequencies in our simulations, theoretical and experimental results. 
As we can obtain that when we slightly change the length $L$, the Fano resonant shapes do appear in the transmission spectrum as shown in Fig. \ref{Fig3} except (b), because Fig. \ref{Fig3} (b) is the BIC case ($L_0 = 85.5$ $\mu m$) which the resonant frequency and corresponding phase of CW are very closed to SRR, as shown in Fig. \ref{Fig2}. 



\begin{figure}[htbp]
	\centering
		\includegraphics[width=0.45\textwidth]{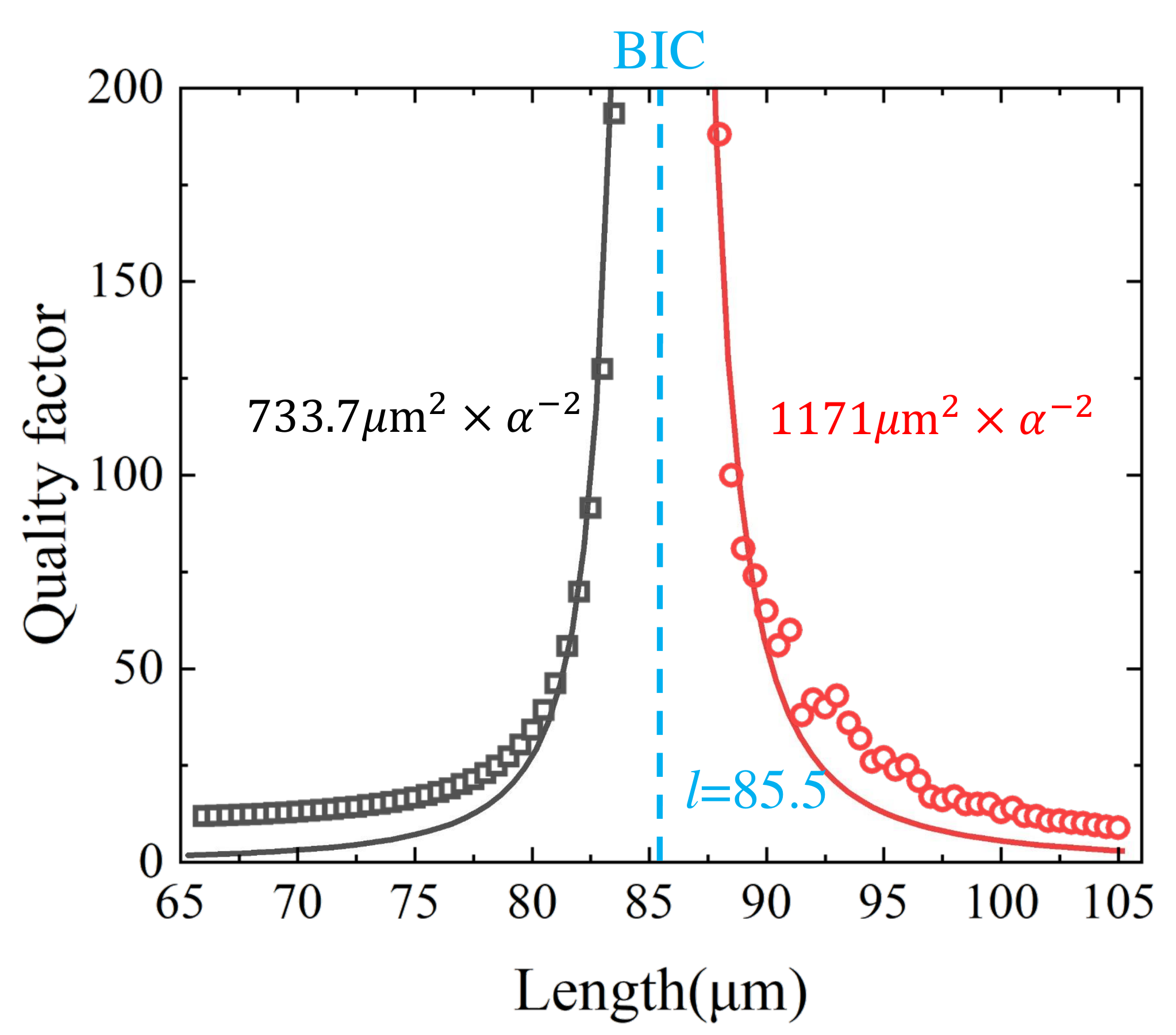}
	\caption{(a)The Q-values of quasi-BIC with two different structures vary the length of CW $L$ from 66 $\mu m$ to 107 $\mu m$. The red and black line are the fitting function of $ Q \propto \alpha^{-2}$, where $\alpha$ is the asymmetric parameter.}
	\label{Fig4}
\end{figure}

\section{Discussion}
From the results of Fig. \ref{Fig3}, our experimental results well satisfied with full-wave simulation and theoretical results. Therefore, we can conclude that our simulated and theoretical results have very good confidence and we can employ our simulated and theoretical results to obtain more features of our device.

In order to better demonstrate the BIC with different structures, we vary the length of CW $L$ from 66 $\mu m$ to 107 $\mu m$, which crosses the same phase and resonant frequency of CW and SRR ($L_0$ = 85.5 $\mu m$). Therefore, from the universal coupled mode theory, we can easily predict that the highest Q-value appears when the same phase and resonant frequency of CW and SRR ,such as the $L_0$ = 85.5 $\mu m$. Therefore, when the length of CW is $L_0$ = 85.5 $\mu m$, it is the BIC point. When the length $L$ of CW is farther away from the BIC point, the Q-value of resonant will be smaller instead. The relationship between the Q-values and the length $L$ of CW can be fitted by function $ Q \propto \alpha^{-2}$, where $\alpha$ is the asymmetric parameter which is given by $\alpha = \frac{L - L_0}{L_0}$ \cite{Azzam2018}. However, in our new type of BIC, the asymmetric parameter $\alpha$ is described by asymmetry between CW and SRR, which means the resonant frequencies between CW and SRR. 
As we can see from the Fig. \ref{Fig4}, it is very easy to see that the Q-values of quasi-BIC are higher, when the length of CW $L$ is more closer to 85.5 $\mu m$ (BIC point) and the Q-values can be fitted by the asymmetric parameter between CW and SRR with the fitting function $ Q \propto \alpha^{-2}$. The fitting functions are given in black line and red line in Fig. \ref{Fig4}. Our results of Fig. \ref{Fig4}  is consistent with our theory and analysis. Hence, our theory can predict and have a very good fitting with BIC or quasi-BIC with CW and SRR, which is the entirely new metamaterial structure for BIC. 

It is worth emphasizing that we employ the CW and SRR as the heterogeneous example to construct BIC and quasi-BIC in this paper, but our conclusions and results can be extended to arbitrary heterogeneous structure to build up the BIC and quasi-BIC, not only for CW and SRR. 

\section{Conclusion}
In this paper, we propose the new configuration of metamaterial BIC with two heterogeneous structures (Cut Wire and Split-Ring Resonator). We firstly theoretically and experimentally demonstrate the BIC and quasi-BIC with two heterogeneous structures. Our theory implies that resonant frequencies and corresponding phases of metamaterial structures are more fundamental physical parameters of BIC and we give the simulation and experimental results to proof our idea. This finding will extend the boundary of metamaterial BIC and could be widely used in applications. 

\section*{Acknowledgements}
This work acknowledges funding from National Science and Technology
Major Project (grant no: 2017ZX02101007-003); National
Natural Science Foundation of China (grant no: 61565004;
61965005; 61975038; 62005059). The Science and
Technology Program of Guangxi Province (grant no:
2018AD19058). Innovation Project of GUET Graduate 
Education(grant no: 2021YCXS129). W.H. acknowledges funding from Guangxi
oversea 100 talent project; W.Z. acknowledges funding from
Guangxi distinguished expert project.

\end{document}